\documentclass[twocolumn,showpacs,prb,amsfonts,amsmath,amssymb,floatfix,groupedaddress]{revtex4-2}
\usepackage{color}
\usepackage{hhline}
\usepackage{mathrsfs}
\usepackage{graphicx}
\usepackage{dcolumn}
\usepackage{bm}
\usepackage{multirow}
\usepackage{booktabs}
\usepackage{tabularx}
\usepackage{afterpage}
\usepackage{amsmath}
\usepackage{algorithm}
\usepackage{algorithmic}
\newcommand{\wmax}{\omega_{\rm max}}

\begin{document}

\title{Uniform electron benchmark for the first-principles $GW_{0}$-Eliashberg theory}

\author{Ryosuke Akashi$^{1}$}
\thanks{akashi.ryosuke@qst.go.jp}
\affiliation{$^1$ National Institutes for Quantum Science and Technology (QST), Ookayama, Meguro-ku, Tokyo 152-8550, Japan}

\author{Hiroshi Shinaoka$^{2}$}
\thanks{shinaoka@mail.saitama-u.ac.jp}
\affiliation{$^2$Department of Physics, Saitama University, Saitama 338-8570, Japan}

\date{\today}
\begin{abstract}
We investigate the numerical behavior of the Eliashberg equations for phonon-mediated 
superconductivity, incorporating normal-state self-energy calculations within the 
consistent $GW_{0}$ approximation. 
We account for the full wavenumber and frequency dependences of both the screened 
Coulomb interaction and phonon-mediated attraction. 
We present results for the prototypical uniform electron gas system with model 
Einstein phonons at temperatures of a few kelvin. 
At extremely low temperatures, we efficiently execute the required convolutions of 
Green's functions and interactions in Matsubara frequency and wavenumber using 
intermediate representation and Fourier convolution techniques. 
In particular, we elucidate the interplay between electron-phonon $\omega$-mass and 
$k$-mass renormalizations of the electronic self-energy in determining the 
normal-state effective mass, spectral weight and the superconducting transition temperature. The electron density regimes where the plasmon effect enhances or suppresses the phonon-mediated superconductivity on top of the static Coulomb effect are revealed.
We compare our comprehensive Eliashberg calculation results with those from 
density functional theory for superconductors, where the functionals have been 
constructed with reference to Eliashberg theory. 
Our model, methods, and results provide a valuable benchmark for first-principles 
superconducting calculations that treat screened Coulomb interaction effects 
non-empirically.
\end{abstract}

\maketitle

\section{Introduction}
First-principles calculation of superconducting transition temperature has long been a challenging goal to conquer for materials science, toward the human dream of computational design of high-temperature superconductors. For phonon-mediated superconducting systems, the Eliashberg theory~\cite{Eliashberg1960,Migdal1958,Scalapino-bookchap,Schrieffer-book} has shown remarkable progress for this with the help of advanced computer resources and algorithms, culminating in accurate non-empirical $T_{\rm c}$ calculations for near-room temperature hydride superconductors at megabar pressures~\cite{sano_effect_2016,sanna_ab_2018, errea_quantum_2020,lucrezi_full-bandwidth_2024}. The key to the success was that the screened Coulomb interaction, a vital factor for determining the absolute value of $T_{\rm c}$, has been made non-empirical than the empirical pseudopotential approach~\cite{morel_calculation_1962,carbotte_properties_1990,aperis_multiband_2018}.

At this stage, the non-empirical screened Coulomb interaction is typically treated with a crude approximation that it should act between electrons instantaneously~\cite{Schrieffer-book,sanna_ab_2018}, as it propagates presumably far faster than the phonon-mediated interaction. This approximation, called {\it static} approximation, has fortunately been corroborated to be accurate for many systems through practices~\cite{sano_effect_2016,sanna_ab_2018,errea_quantum_2020,wang_efficient_2020,davydov_ab_2020,pellegrini_eliashberg_2022,lucrezi_full-bandwidth_2024,pellegrini_ab_2024}. However, this approach misses the dynamical effect of the screened Coulomb interaction, that is represented by the frequency dependence of the dielectric function in the energy scale of the normal electrons.

The frequency-dependent screened Coulomb interaction in the Eliashberg theory has been extensively studied for over five decades. The early studies mainly focused on the uniform electron gas, with particular interest in whether the screened Coulomb interaction can induce superconductivity without phonons~\cite{takada_plasmon_1978,rietschel_role_1983,takada_plasmon_1992,takada_s_1993}. 
The quantitative effect of the frequency dependence is of broader interest in systems where it coexists with the phonon-mediated interaction~\cite{koonce_superconducting_1967,richardson_effective_1997, takada_theory_2011,akashi_development_2013,akashi_density_2014,christiansson_superconductivity_2022,int_veld_screening_2023}. Even if it cannot induce superconductivity by itself, it can still alleviate the negative effect of the static component of the Coulomb interaction for $T_{\rm c}$. However, first-principles inclusion of this in the Eliashberg theory poses formidable challenges to researchers. Conceptually, we have to enter the regime where the neglect of the higher-order vertices~\cite{takada_insignificance_1992} is not justified. Even if we stop at the lowest-order $GW$ level for the self-energy~\cite{hedin_new_1965}, we have to achieve a compatible accuracy of the self-energy calculation in the normal electron energy scale and the superconducting gap calculation in the scale of a few Kelvin. Some early studies managed this by careful numerical extrapolation to extremely low temperature~\cite{richardson_effective_1997,wang_origin_2023}. However, systematic methods that solve this multiple energy-scale problem are still lacking. 

In this paper, we devise a computationally efficient method to solve the Eliashberg equations with full frequency and wavenumber dependence of the screened Coulomb interaction and phonon-mediated attraction. The electron self-energy is calculated at ``$GW_{0}$" level~\cite{Aryasetiawan-review}, where the screened Coulomb and phonon-mediated interactions ($W_{0}$) are not recalculated on the self-consistent update of the electron Green function, in line with the usual practice of the standard Eliashberg calculation. We apply the method to a model with uniform electrons and Einstein-type phonon spectrum, which accurately captures essential effects of the frequency and wavenumber dependence of the interactions. The screened Coulomb interaction is calculated with the random phase approximation~\cite{Pines-RPA}, whereas the phonon-mediated attraction is modeled with a simplified Einstein spectrum with a wavenumber cutoff. 

The self-energy calculation and Eliashberg equation solution require us to evaluate polynomially decaying integrals over the Matsubara frequency and wavenumber.
For the former, we use the expansion by the intermediate representation (IR) basis~\cite{shinaoka_compressing_2017,Shinaoka2022-xv}, which reduces the number of quadrature points from $O(1/T)$ to $O(\log(1/T))$ at temperature $T$~\cite{Li2020-kb}.
For the latter, we develop an algorithm based on the Fourier convolution theorem, which also enables us to reduce the number of internal integral dimensions. Using the developed method, we reveal unexplored effects of the {\it shift} term: In the first-principles Eliashberg theory, the shift term or wavenumber dependence of the self-energy is legitimately neglected with the static Coulomb approximation~\cite{Schrieffer-book} but otherwise requires tedious, slowly converging integrals.
In the first-principles calculations this shift is often assumed to be included in the Kohn-Sham~\cite{hohenberg_inhomogeneous_1964,kohn_self-consistent_1965} basis states~\cite{davydov_ab_2020,pellegrini_ab_2024} but this assumption obviously fails in the uniform electron gas~\cite{akashi_revisiting_2022}.
We establish how the shift term affects the normal-state effective mass and superconducting transition temperature ($T_{\rm c}$) quantitatively in the uniform limit, toward which the RPA-based first-principles methods for superconductors should be benchmarked. To demonstrate the usefulness of the current results, we compare $T_{\rm c}$s calculated by the density functional theory for superconductors~\cite{oliveira_density-functional_1988,luders_ab_2005,marques_ab_2005}, whose exchange-correlation functionals have been constructed to accurately reproduce the Eliashberg results with reduced computational cost. 

\section{Theory and setup}
Throughout the paper, we adopt the atomic unit unless specified otherwise. The Eliashberg theory is a perturbation theory for the electron-phonon coupled system, in which we work on a set of equations for the Green's function ${\bf G}$
\begin{eqnarray}
  \!\!{\bm \Sigma}({\bm r}_{1},\! {\bm r}_{2},i\omega_{j})
  &=&\!
  -T\sum_{\omega_{j'}}\!\!
  \sigma_{3}
  {\bm G}({\bm r}_{1},\! {\bm r}_{2},i\omega_{j'})
  \sigma_{3}
  W({\bm r}_{1},\! {\bm r}_{2},i(\omega_{j}\!-\!\omega_{j'}\!))
  \nonumber \\
 &&-V_{\rm eff}({\bm r}_{1}, {\bm r}_{2})\sigma_{3},
  \label{eq:Eliashberg-eq}
  \\
  {\bm G}^{-1}&=&{\bm G}_{0}^{-1}-{\bm \Sigma}
  \label{eq:Dyson}
  .
\end{eqnarray}
Here and hereafter, the bold letters denote the quantities in the Nambu 2$\times$2 notation~\cite{nambu_quasi-particles_1960,Gorkov1958}. The interaction $W$ is composed of the phonon mediated attraction and screened Coulomb interaction: $W=W_{\rm ph}+W_{\rm C}$. ${\bf G}_{0}=\left((G_{0}, 0), (0, G_{0}^{\ast})\right)$ denotes the non-interacting Green's function, which is related to a one-body effective mean-field potential $V_{\rm eff}$ by $\left[i\omega-\left(-\hat{\nabla}^2/2\right)-\hat{V}_{\rm H}-\hat{V}_{\rm eff}+\mu_{0}\right]G_{0}=\delta({\bf r}-{\bf r}')$. Here $\hat{V}_{\rm H}$ is composed of the ionic and electronic Hartree potentials, whereas $V_{\rm eff}$ can be taken arbitrarily. A practical choice is the Kohn-Sham exchange-correlation potential~\cite{kohn_self-consistent_1965}, by which $\{i\}$ corresponds to the Kohn-Sham basis.

Inserting Eq.~(\ref{eq:Dyson}) into Eq.~(\ref{eq:Eliashberg-eq}), decomposing the terms into the Pauli matrices $(\sigma_{0},\sigma_{1}, \sigma_{2}, \sigma_{3})$ and assuming that ${\bf G}$ and ${\bf \Sigma}$ are diagonalized by a basis set common to ${\bf G}_{0}$, We get the celebrated expression of the Eliashberg equations 
\begin{eqnarray}
  Z_{i}(i\omega_{j})
  &=&
  1
  -
  T
  \sum_{i'j'}
  \frac{\omega_{j'}}{\omega_{j}}\frac{Z_{i'}(i\omega_{j'})}{\Theta_{i'}(i\omega_{j'})}
  W_{ii'}(i(\omega_{j}-\omega_{j'}))
  ,
  \label{eq:Eliashberg-Z}
  \\
  \chi_{i}(i\omega_{j})
  &=&
  T
  \sum_{i'j'}
  \frac{\xi_{i'}+\chi_{i'}(i\omega_{j'})}{\Theta_{i'}(i\omega_{j'})}
  W_{ii'}(i(\omega_{j}-\omega_{j'}))
  \nonumber \\
  && \hspace{50pt} -V_{{\rm eff},i}
  ,
  \label{eq:Eliashberg-chi}
  \\
  \phi_{i}(i\omega_{j})
  &=&
  -
  T
  \sum_{i'j'}
  \frac{\phi_{i'}(i\omega_{j'})}{\Theta_{i'}(i\omega_{j'})}
  W_{ii'}(i(\omega_{j}-\omega_{j'}))
  ,
  \label{eq:Eliashberg-phi}
  \\
  \Theta_{i}(i\omega_{j})
  &=&
  \left[
  \omega_{j}
  Z_{i}(i\omega_{j})
  \right]^2
  \!+\!
  \left[
  \xi_{i}
  +
  \chi_{i}(i\omega_{j})
  \right]^2
  \!+\!
  \left[
  \phi_{i}(i\omega_{j})
  \right]^2
  \nonumber \\    
\end{eqnarray}
with the particle-number conservation condition that determines the chemical potential $\mu_{0}$.
\begin{eqnarray}
  N=\sum_{ij}\frac{i\omega_{j}Z_{i}(i\omega_{j})+\xi_{i}+\chi_{i}(i\omega_{j})}{\Theta_{i}(i\omega_{j})}e^{i\omega_{j}0^{+}}
  .
  \label{eq:N-conserve}
\end{eqnarray}
$\xi_{i}$ is the single-particle energy eigenvalue related to the diagonalized $G_{0ii}\equiv G_{0i}$ by $G_{0i}(i\omega_{j})=\left(i\omega_{j}-\xi_{i}\right)^{-1}$, $V_{{\rm eff},i}\equiv\langle i|V_{\rm eff}|i \rangle$ is the matrix element with respect to the basis $i$, and $W_{ii'}$ denotes the matrix element of the interaction for the Cooper pairs $\langle ii^{\ast}|W|i'i'^{\ast}\rangle$.

By linearizing the equations with respect to the anomalous component $\phi$, the self consistent solution of Eqs.~(\ref{eq:Eliashberg-Z})--(\ref{eq:Eliashberg-phi}) is simplified to sequential calculations of the normal-state self-energy
\begin{eqnarray}
  \Sigma_{i}(i\omega_{j})&=&-T\sum_{i'j'}G_{i'}(i\omega_{j'})W_{ii'}(i(\omega_{j}-\omega_{j'})),
    \label{eq:selfene_normal}
  \\
  \frac{1}{G_{i}(i\omega_{j'})}&=&
  \frac{1}{G^{(0)}_{i}(i\omega_{j'})}-\Sigma_{i}(i\omega_{j})
\end{eqnarray}
and the anomalous part
\begin{eqnarray}
  \phi_{i}(i\omega_{j})
  \!&=&\!-T\sum_{i'j'}\frac{\phi_{i'}(i\omega_{j'})W_{ii'}(i(\omega_{j}-\omega_{j'}))}{\bar{\Theta}_{i}(i\omega_{j})}
  , \label{eq:linear-Eliashberg}  \\
  \bar{\Theta}_{i}(i\omega_{j})&=&\left[\omega_{j'}Z_{i'}(i\omega_{j'})\right]^2\!+\!\left[\xi_{i'}+\chi_{i'}(i\omega_{j'})\right]^2.\nonumber
\end{eqnarray}
The latter has a nontrivial solution only at the transition temperature.
Note that $Z_{i}$ and $\chi_{i}$ in Eq.~(\ref{eq:linear-Eliashberg}) are the normal state values related to $\Sigma_{i}$ in Eq.~(\ref{eq:selfene_normal}) by
\begin{eqnarray}
  \hspace{-5pt}  i\omega_{j}Z_{i}(i\omega_{j})&=&\frac{1}{2}\left[\Sigma_{i}(i\omega_{j})-\Sigma_{i}(-i\omega_{j})\right]\equiv \Sigma^{(Z)}_{i}(i\omega_{j}), \label{eq:selfene_antisym}\\
    \chi_{i}(i\omega_{j})&=&\frac{1}{2}\left[\Sigma_{i}(i\omega_{j})+\Sigma_{i}(-i\omega_{j})\right].
    \label{eq:selfene_sym}
\end{eqnarray}
The factor $Z_{i}$, depending quadratically on $\omega$ and $\xi_{j}$ in the vicinity of the Fermi surface, yields the renormalization of the quasiparticle Green function as $G_{i}(i\omega_{j})\sim 1/[1+Z_{i}(0)]/\left\{i\omega_{j}-\xi_{i}/[1+Z_{i}(0)]\right\}$.
The factor $\chi_{i}$, which depend linearly on $\xi_{i}$ and quadratically on $\omega$, induces a shift $\xi_{j}\rightarrow\xi_{j}+\chi_{j}(0)$ around the Fermi level.

As we work on the uniform electron gas throughout this paper, the basis $\{i\}$ is set to be the plane waves labeled by wavenumber ${\bf k}$. We set the non-interacting Green's function $G_{0}$ to be the free-electron one $(\xi_{i}\rightarrow \xi({\bf k})=k^2/2-\mu_{0})$ and therefore the effective potential $V_{\rm eff}=0$. The Wigner-Seitz radius $r_{\rm s}$ is the sole tunable parameter for the electronic system. The Fermi wavenumber $k_{\rm F}=\left(9\pi/4\right)^{1/3}/r_{\rm s}$ is not changed by the interaction, with which we do not have to treat the particle-number conservation (Eq.~(\ref{eq:N-conserve})) explicitly. The interacion $W_{ii'}(i\nu)\equiv W({\bf k}_{i}-{\bf k}_{i'},i\nu)$ is composed of the screened Coulomb and phonon-mediated contributions:
\begin{equation}
W({\bf k}_{i}-{\bf k}_{i'},i\nu)=W_{\rm C}({\bf k}_{i}-{\bf k}_{i'},i\nu)+W_{\rm ph}({\bf k}_{i}-{\bf k}_{i'},i\nu).\label{eq:Wdecomp}
\end{equation}
The screened Coulomb interaction is given by 
\begin{eqnarray}
  W_{\rm C}({\bf k},i\nu)&=&\frac{V({\bf k})}{\varepsilon_{\rm RPA}({\bf k}, i\nu)}
  ,
\end{eqnarray}
with $\varepsilon_{\rm RPA}$ being the Lindhard dielectric function~\cite{Lindhard1954} and $V({\bf k})$ being the bare Coulomb interaction $4\pi/k^2$.
The phonon-mediated interaction is given by
\begin{eqnarray}
 W_{\rm ph} ({\bf k},i\nu)
 =-\sum_{l}|g_{l}({\bf k})|^{2}\frac{2\omega_{l}({\bf k})}{\nu^2 + \omega_{l}({\bf k})},
\end{eqnarray}
where $l$ labels the phonon modes. We adopt an Einstein model spectrum $\omega_{l}({\bf k})=\omega_{\rm E}$ with Debye wavenumber cutoff $k_{\rm D}$ and model matrix element
\begin{eqnarray}
  \sum_{l}|g_{l}({\bf k})|^{2}=A\theta(k_{\rm D}-k),\nonumber \\
A=\lambda\frac{\pi^2 \omega_{\rm E}}{k_{\rm F}}\times {\rm max}\left[1,\left(2k_{\rm F}/k_{\rm D}\right)^2\right]
\end{eqnarray}
so that the Fermi surface average of the interaction amounts to $\lambda$
\begin{eqnarray}
   &&\frac{2}{N(0)}\int\frac{d^3 k}{(2\pi)^3}\int\frac{d^3 k'}{(2\pi)^3}W_{\rm ph}({\bf k}-{\bf k}',0)\delta(\xi_{k})\delta(\xi_{k'})\nonumber \\
   &&\hspace{50pt}=\lambda 
   .
\end{eqnarray}
Factor $N(0)$ denotes the electronic density of state at the Fermi level. Thus, for a given $r_{\rm s}$, $W^{\rm ph}$ is controlled by $\lambda, \omega_{\rm E}$, and $ k_{\rm D}$.

\section{Numerics}
The linearized Eliashberg equation~(\ref{eq:linear-Eliashberg}) in the uniform case has the structure
\begin{eqnarray}
  \phi({\bf k},i\omega)
  \!=\!
  \sum_{{\bf k}',\omega'} K({\bf k}-{\bf k}',i(\omega-\omega');T)\frac{\phi({\bf k}',i\omega')}{\bar{\Theta}({\bf k}',i\omega')}
  ,
  \label{eq:linear-Eliashberg-unif}
\end{eqnarray}
where $K$ represents an abstract kernel of the equation. As the system is uniform, we may assert $\phi({\bf k},i\omega)=\phi_{l}(k,i\omega)Y_{lm}(\hat{\bf k})$ with $Y_{lm}$ and $\hat{\bf k}$ being the spherical harmonics and solid angle, respectively, and Eq.~(\ref{eq:linear-Eliashberg-unif}) is transformed to
\begin{eqnarray}
  \phi_{l}(k,i\omega)
  =
  \sum_{k',\omega'} K_{l}(k,k',i(\omega-\omega');T)\frac{\phi_{l}(k',i\omega')}{\bar{\Theta}(k',i\omega')}
  .
\end{eqnarray}
Although we do not discuss a specific form of $K_{l}$ above, to derive this, an integral is taken with respect to the polar angle between ${\bf k}$ and ${\bf k}'$~\cite{Schrieffer-book, Allen-Mitrovic}. The straightforward approach to find $T_{\rm c}$ is (i) calculate the matrix $K_{l}(k,k',i(\omega-\omega');T)$ and next (ii) calculate the signed largest eigenvalue of $\frac{K_{l}(k,k',i(\omega-\omega');T)}{\Theta(k',i\omega')}$ to see if it amounts to unity. These steps respectively require the computational costs of $O(N_{k}^2 N_{\omega}^2 N_{\theta})$ and $O(N_{k}^2 N_{\omega}^2)$, where $N_{k}$, $N_{\omega}$ and $N_{\theta}$ are the numbers of quadrature points for $k$, $\omega$ and polar angle integrals. 

For the convolution with respect to $\omega'$, Ref.~\cite{sano_effect_2016} proposed to use the Fourier convolution technique
\begin{eqnarray}
\sum_{\omega'}f(\omega-\omega')g(\omega')\nonumber=\mathcal{F}_{\tau\rightarrow\omega}\left[\mathcal{F}^{-1}_{\omega\rightarrow\tau}(f)\mathcal{F}^{-1}_{\omega\rightarrow\tau}(g)\right]
,
\label{eq:Fconv-w}
\end{eqnarray}
and implemented this with the Fast Fourier Transformation (FFT) algorithm.
However, for a required accurary $\epsilon$, $N_{\omega}$ scales as $N_{\omega}\propto O(\epsilon^{-1} \wmax/T)$, where $\wmax$ is a ultra violet cutoff for the real frequency (e.g., band width).
Thus, $N_{\omega}$ becomes prohibitively large for the current superconducting calculations, where the grid cover from a few Hartree to a few Kelvin.

To overcome this issue, we use the convolution technique proposed in Ref.~\cite{wang_efficient_2020}.
This approach employs the sparse frequency and time grids~\cite{Li2020-kb} generated based on the IR~\cite{shinaoka_compressing_2017} (see also Ref.~\cite{Shinaoka2022-xv} for a review).
In this approach, $N_{\omega}$ grows as $N_{\omega} \propto \log (1/\epsilon)\log(\wmax/T)$, only logarithmically for $\epsilon$ and $\wmax$.
We do not utilize FFT for the transforms in Eq.~(\ref{eq:Fconv-w}) then but still the IR method is advantageous because of the drastically reduced $N_{\omega}$.
In practice, $N_{\omega}\simeq 170$ suffices for the current calculations below 1~K.

Also, we apply the Fourier convolution technique to the ${\bf k}'$-convolution:
\begin{eqnarray}
  \sum_{{\bf k}'}f({\bf k}-{\bf k}')g({\bf k}')
  =
  \mathcal{F}_{{\bf r}\rightarrow{\bf k}}\left[\mathcal{F}^{-1}_{{\bf k}\rightarrow{\bf r}}(f)\mathcal{F}^{-1}_{{\bf k}\rightarrow{\bf r}}(g)\right]
  .
\end{eqnarray}
For the spherically symmetric systems in three dimensions, $\mathcal{F}^{-1}_{{\bf k}\rightarrow{\bf r}}$ reads as the one-dimensional spherical Bessel transform
\begin{eqnarray}
  \mathcal{F}^{-1}_{{\bf k}\rightarrow{\bf r}}(f_{l})= f_{l}(r)=\frac{i^l}{2\pi^2}\int_{0}^{\infty} dk k^2j_{l}(kr)f_{l}(k)
  ,
  \label{eq:SBT}
\end{eqnarray}
where $f({\bf k})=\sum_{lm}Y_{lm}(\hat{\bf k})f_{l}(k)$ and we dropped the axial angle index $m$ as trivial. We do not use the FFT as we need to treat extremely non-uniform $k$ points for efficiently capturing the sensitive $k$ dependence of the integrands. With this procedure we do not need an integral for the polar angle between ${\bf k}$ and ${\bf k}'$.

We thus recast the original eigensolution problem to the sequential Fourier transforms by cost $O(N_{k}^{2}N_{\omega})+O(N_{k}N_{\omega}^{2})$ (the number of points in $r$ is assumed to be $\sim N_{k}$), where the vector dimensions of the quantities stored during the calculations are utmost $O(N_{k}N_{\omega})$ and $N_{\omega}\propto {\rm log}(1/T)$.
A similar advantage may be taken for non-uniform systems where $\phi$ becomes dependent on two wavenumbers, which we will attempt in future studies.

On evaluating the spherical Bessel transform Eq.~(\ref{eq:SBT}), we need to manage integrals of slowly decaying and oscillating functions. For this purpose, we utilize the continuous Euler transformation proposed by Ooura~\cite{ooura_continuous_2001}
\begin{eqnarray}
  \int_{0}^{\infty} dx e^{ikx}f(x) \approx\int_{0}^{x_{\rm max}} dr w(x;p,q)e^{ikx}f(x),
  \label{eq:contEuler}
\end{eqnarray}
where $w(x;p,q)=\frac{1}{2}{\rm erfc}\left(x/p\ -q\right)$.
This method seems to introduce a smooth cutoff, as broadly practiced for integrals involving the Coulomb interaction~\cite{Alavi_PRB2008}, but with appropriate constraints on parameters $p$ and $q$, the right-hand side of Eq.~(\ref{eq:contEuler}) converges to the exact value with error $O(e^{-q^2})$ for a broad class of functions $f$. An empirically efficient setting $(p,q)=(\sqrt{x_{\rm max}},\sqrt{x_{\rm max}}/2)$ is used throughout this work. 

The overall workflows for normal and anomalous self-energies are summarized as Algorithms \ref{alg:self_energy} and \ref{alg:lin_Eliashberg}. For the Fourier transforms between $\omega$ and $\tau$ spaces, we use the IR~\cite{shinaoka_compressing_2017} as performed in Ref.~\cite{wang_efficient_2020}. We applied the aforementioned continuous Euler transform to SBTs from $r$ to $k$ spaces in lines \ref{lin:SBT1}, \ref{lin:SBT2} in Algorithm \ref{alg:self_energy} and lines \ref{lin:SBT3} and \ref{lin:SBT4} in Algorithm \ref{alg:lin_Eliashberg}. To evaluate $T_{\rm c}$, those two Algorithms are enclosed in the loop for temperature update, at the beginning of which the IR basis is numerically generated and stored for each temperature. Algorithms concerning IR basis were executed with \texttt{SparseIR.jl}~\cite{Wallerberger2023-yq} in Julia. See the footnote for detailed settings~\footnote{We generated basis with the parameters $\omega_{\rm max}=20$ and accuracy $\epsilon=5\times 10^{-13}$. When solving the linearized Eliashberg equation, we trimmed components of $\phi$ in the IR with index larger than $\frac{2}{3}N_{l{\rm max}}$ ($N_{l{\rm max}}$: Number of IR basis) at each iteration}.

\begin{algorithm}[H]
  \caption{Calculate normal self-energy}
  \label{alg:self_energy}
  \begin{algorithmic}[1]
    \STATE Calculate $W(k,i\omega)-V(k)$ 
    \STATE FT from $W(k,i\omega)-V(k)$ to $W(k,i\tau)-V(k)$
    \STATE SBT from $W(k,i\tau)-V(k)$ to $W(r,i\tau)-V(r)$
    \STATE Initialize $G(k,i\omega)$ by $G_{0}(k,i\omega)$
    \REPEAT
      \STATE FT from $G(k,i\omega)$ to $G(k,i\tau)$
      \STATE SBT from $G(k,i\tau)$ to $G(r,i\tau)$
      \STATE Elementwise product. $[GV](r,0)$ and $[G(W-V)](r,i\tau)$ 
      \STATE SBT from $[GV](r,0)$ to $[GV](k,0)=-\Sigma_{x}(k)$ \label{lin:SBT1}
      \STATE SBT from $[G(W-V)](r,\tau)$ to $[G(W-V)](k,\tau)$ \label{lin:SBT2}
      \STATE FT from $[G(W-V)](k,\tau)$ to $[G(W-V)](k,\omega)$
      \STATE $\Sigma(k,i\omega)=-[G(W-V)](k,\omega)+\Sigma(k)$
      \STATE Dyson equation. $G=\left[1-G_{0}\Sigma\right]^{-1}G_{0}$
    \UNTIL{$\Sigma$ is converged}
  \end{algorithmic}
\end{algorithm}

\begin{algorithm}[H]
  \caption{Iteratively solve the linearized Eliashberg equation. $\Sigma(k,i\omega)$ and $W(r,i\tau)-V(r)$ are stored already.}
  \label{alg:lin_Eliashberg}
  \begin{algorithmic}[1]
    \STATE Calculate $\Theta(k,i\omega)$
    \STATE Initialize $\phi(k,i\omega)$
    \REPEAT
      \STATE FT from $\frac{\phi}{\Theta}(k,i\omega)$ to $\frac{\phi}{\Theta}(k,i\tau)$
      \STATE SBT from $\frac{\phi}{\Theta}(k,i\tau)$ to $\frac{\phi}{\Theta}(r,i\tau)$
      \STATE Elementwise product. $[\frac{\phi}{\Theta}V](r,0)$ and $[\frac{\phi}{\Theta}(W-V)](r,i\tau)$ 
      \STATE SBT from $[\frac{\phi}{\Theta}V](r,0)$ to $[\frac{\phi}{\Theta}V](k,0)$ \label{lin:SBT3}
      \STATE SBT from $[\frac{\phi}{\Theta}(W-V)](r,\tau)$ to $[\frac{\phi}{\Theta}(W-V)](k,\tau)$ \label{lin:SBT4}
      \STATE FT from $[\frac{\phi}{\Theta}(W-V)](k,\tau)$ to $[\frac{\phi}{\Theta}(W-V)](k,\omega)$
      \STATE $\delta \phi=\frac{\phi}{\Theta}(W-V)](k,\omega)+[\frac{\phi}{\Theta}V](k,0)$
      \STATE $\phi'(k,i\omega)=\phi(k,i\omega)-a\delta \phi(k,i\omega)$
      \STATE $\varepsilon=\frac{1}{a}\left[1-||\phi'||/||\phi||\right]$
      \STATE $\phi\leftarrow\phi'$
    \UNTIL{$\varepsilon$ is converged}
  \end{algorithmic}
\end{algorithm}

The functions involved have ``hot spots" in $k$ space: $k\simeq 0$ from the long-range nature of the Coulomb interaction; $k\simeq k_{\rm F}$ reflecting the Fermi surface and divergent derivative in the Coulomb self-energy; $k\simeq k_{\rm D}$ for the cutoff of the phonon-mediated interaction; $k\simeq 2k_{\rm F}$ for the singularity in the RPA dielectric function. We generated non-uniformly spaced $k$ points that accumulate around those spots, using the double exponential quadrature~\footnote{
The double exponential (DE) quadrature gives the quadrature points and weights in the range $[x_{\rm max},x_{\rm min}]$ as
\begin{eqnarray}
  x_{j} &=& x_{0}+W{\rm tanh}\left[\frac{\pi}{2}{\rm sinh}(jh)\right],\\
  w_{j} &=& h\frac{W}{2}\pi{\rm cosh}(jh)\frac{1}{\left[{\rm cosh}\left(\frac{\pi}{2}{\rm sinh}(jh)\right)\right]^2}
\end{eqnarray}
with
\begin{eqnarray}
  x_{0}=\frac{1}{2}(x_{\rm max}+x_{\rm min}), \ \ W=\frac{1}{2}(x_{\rm max}-x_{\rm min})
\end{eqnarray}
so that the integral of a function $f(x)$ over the range $[x_{\rm min},x_{\rm max}]$ is approximated as
\begin{eqnarray}
  \int_{x_{\rm min}}^{x_{\rm max}}dx f(x)\approx \sum_{j}w_{j}f(x_{j})
  .
\end{eqnarray}
We divided the $k$-range $[0,k_{\rm max}]$ into $[0,k_{\rm F}]$, $[k_{\rm F},k_{\rm D}]$, $[k_{\rm D},2k_{\rm F}]$, $[2k_{\rm F},k_{\rm max}]$ and generated $N_{k}$ quadrature points for each range. Setting $h=2.5/N_{k}$, we obtained a 4$N_{k}$ $k$-point set that accumulates around the boundaries. We set $k_{\rm max}=7.0$.
}

\begin{figure}[t!]
  \begin{center}
   \includegraphics[scale=0.37]{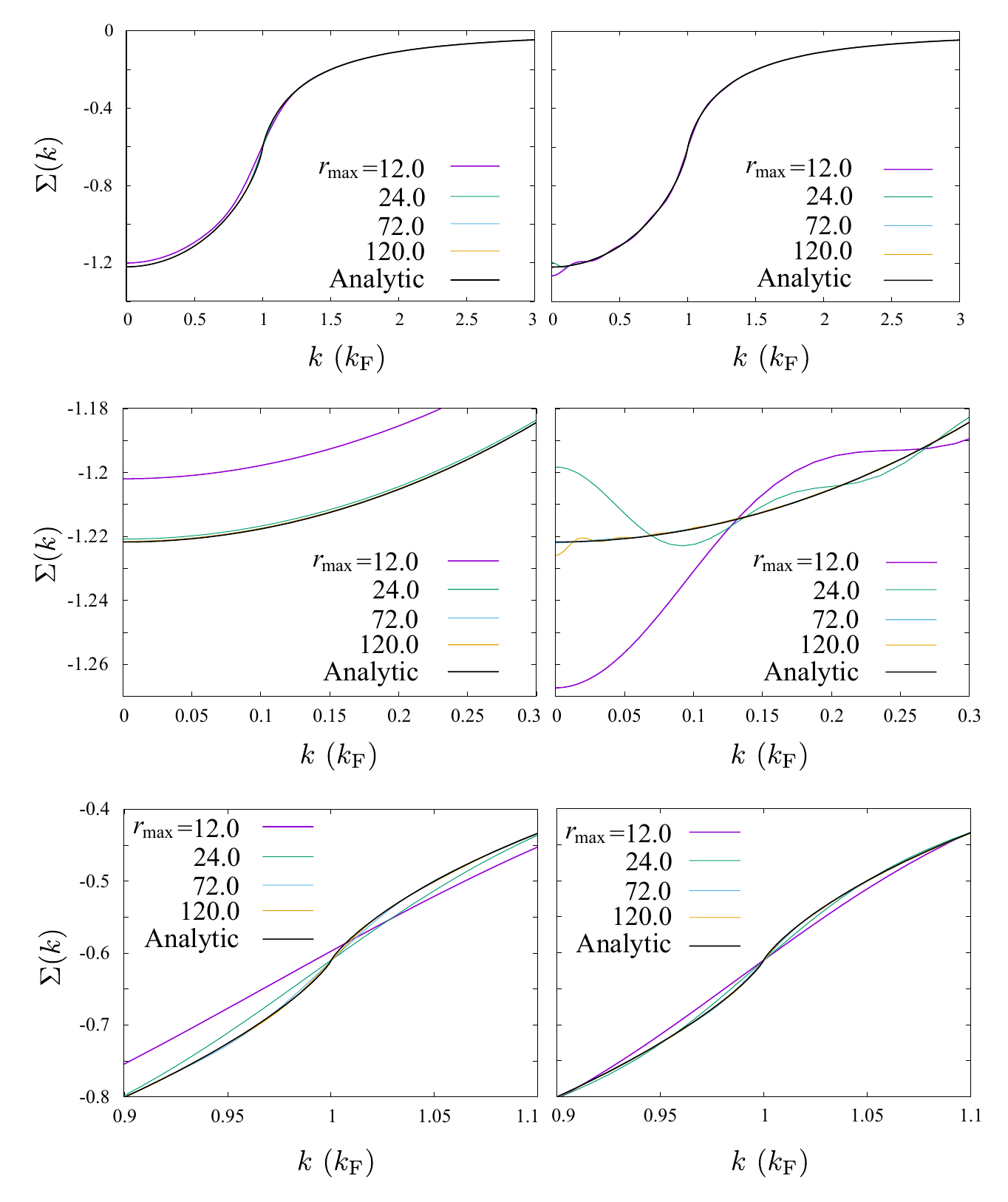}
   \caption{
   Convergence of the bare Fock part of the self-energy Eq.~(\ref{eq:selfene-Fock}) with respect to the cutoff $r_{\rm max}$. (Left) Results with the continuous Euler weight. (Right) Results with the uniform weight. Middle and bottom panels are close-up views at $k\simeq 0$ and $\simeq k_{\rm F}$, respectively.
   } 
   \label{fig:Fock-test}
   \end{center}
 \end{figure}

\subsection*{Test: Bare Fock term}
We examine the efficiency of the continuous Euler transformation with the bare Fock term in the self-energy 
\begin{eqnarray}
  \Sigma(k)&=&-T\sum_{\omega_{j}}\int\frac{d^3 k'}{(2\pi)^3}G({\bf k}-{\bf k}',i\omega_{j})V({\bf k}')
  \nonumber \\
  &\approx&-\int d^3 r w(r) e^{-i{\bf k}\cdot{\bf r}}f_{\beta}(r)V(r)
  ,
  \label{eq:selfene-Fock}
\end{eqnarray}
where $f_{\beta}(r)=\int d^3{k}/(2\pi)^3 e^{i{\bf k}\cdot{\bf r}}f_{\beta}(\xi_{q})$, $V(r)=1/r$ and $w(r)$ is the quadrature weight. The quadrature points for $r$ were taken to be equally spaced with an interval of 0.12.

Figure~\ref{fig:Fock-test} compares the results with the exact analytic formula. Compared with a uniform weight $w(r)=1$, the continuous Euler method ($w(r)=w(r;p,q)$) shows fast and uniform convergence in the overall $k$ regions.
In particular, it converges in a very stable manner at small $k$. The Fock term has the celebrated derivative singularity at $k=k_{\rm F}$. Around this point, the continuous Euler method is not apparently superior, which is perhaps because one needs a large number of sampling periods for evaluating the rapidly varying Fourier components around $k\sim k_{\rm F}$.
The diverging slope, which disappears when considering the screening at $i\omega=0$ but remains at $i\omega\neq 0$, is related to a sharp ``trough"~\cite{takada_plasmon_1992} in Eliashberg's eigenvector $\phi(k,i\omega)$. To assure stable convergence for both regions we set the cutoff $r_{\rm max}=120.0$ hereafter.

\section{Results}

\begin{figure}[t!]
  \begin{center}
   \includegraphics[scale=0.60]{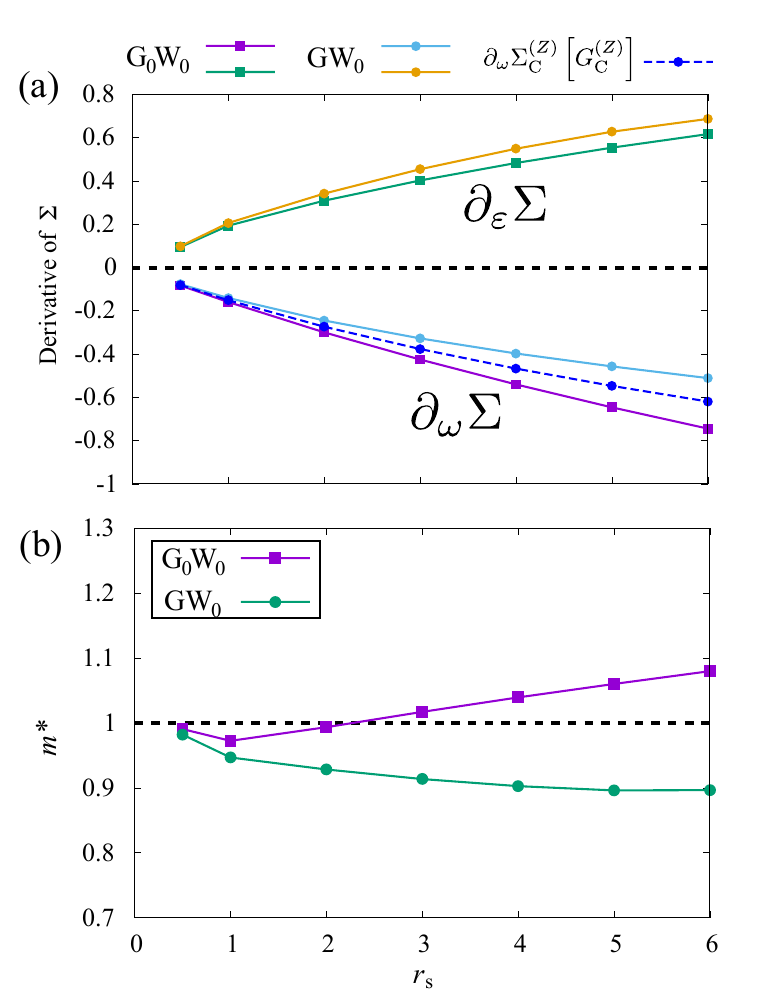}
   \caption{Normal self-energy properties calculated with $W_{\rm ph}=0$. (a) Derivatives of the self-energy evaluated at $(k,i\omega)=(k_{\rm F},0)$. Labels ``$G_{0}W_{0}$'' and ``$GW_{0}$'' denote $\Sigma_{\rm C}\left[G_{0}\right]$ and $\Sigma_{\rm C}\left[G_{\rm C}\right]$ in our specific notation, respectively. (b) Effective mass computed from the electronic contribution of the self-energy $\Sigma_{\rm el}$.} 
   \label{fig:invmass-C-G0-G}
   \end{center}
 \end{figure}

We examined the self-energy in the uniform electron gas plus Einstein phonon system and its impact on effective mass correction and superconducting $T_{\rm c}$. 

\subsubsection*{Notation for different approximations}
In what follows, we discuss results obtained using different approximations to the self-energy functional and to the normal Green function.
The normal self-energy, $\Sigma[G]$, depends on (i) the approximation chosen for the self-energy functional of $G$ and (ii) the approximation used for the normal Green function $G$ that serves as its argument.
Equation \eqref{eq:selfene_normal} decomposes $\Sigma$ into phononic and Coulombic contributions, $\Sigma = \Sigma_{\rm ph} + \Sigma_{\rm C}$.
We indicate the treatment of the Matsubara-frequency dependence via a superscript: $\Sigma^{(Z)}$ denotes that the antisymmetric approximation to $\Sigma$ is used [Eq.~\eqref{eq:selfene_antisym}], whereas the absence of a superscript indicates the full frequency dependence.
For the $G$ in brackets, the superscript and subscript specify the self-energy approximation used to obtain $G$ self-consistently.
Hence, a calculated self-energy is fully specified by the notation $\Sigma^{({\rm A})}_{\rm B}\left[G^{({\rm C})}_{\rm D}\right]$.

For later reference, we define the partial derivatives $\partial/\partial \omega \equiv \partial_{\omega}$ and $\partial/\partial \varepsilon \equiv \left(\partial/\partial k\right)/\left(d\varepsilon(k)/dk\right) \equiv \partial_{\varepsilon}$ with $\varepsilon(k) = k^2/2$.
Unless stated otherwise, these derivatives are evaluated at the Fermi level $(k = k_{\rm F}, i\omega = 0)$.

\subsection{Normal-state effective mass and spectral weight}
The effective mass including $\Sigma_{\rm ph}$ and $\Sigma_{\rm C}$ is determined as follows~\cite{Grimvall-book}:
\begin{eqnarray}
    m^{\ast}=\frac{1-\partial_{\omega} \Sigma_{\rm C}-\partial_{\omega} \Sigma_{\rm ph}}{1+\partial_{\varepsilon} \Sigma_{\rm C}+\partial_{\varepsilon}\Sigma_{\rm ph}}
    .
    \label{eq:mass-tot}
\end{eqnarray}
We remember that the $\omega$-mass terms ($\partial_{\omega}\Sigma$) and $k$-mass terms ($\partial_{\varepsilon}\Sigma$) exclusively originate from the terms antisymmetric and symmetric with respect to $\omega$. Below, we investigate how the total $m^{\ast}$ varies depending on the approximations.

\subsubsection{$W_{\rm ph}=0$ case}
The effective mass without $\Sigma_{\rm ph}$, $m_{\rm el}^{\ast}\equiv \left(1-\partial_{\omega} \Sigma_{\rm C}\right)/\left(1+\partial_{\varepsilon} \Sigma_{\rm C}\right)$, with the RPA screened $W_{\rm C}$ is an established quantity ~\cite{hedin_new_1965,rietschel_role_1983}. Nevertheless, the analysis including the phononic contribution and decomposition into the $Z$ and $\chi$ parts, which becomes particularly relevant in the Eliashberg context, provides novel insight into how the total $m^{\ast}$ is determined. Note that in this case, $r_{\rm s}$ is the only parameter controlling the system.

In Fig.~\ref{fig:invmass-C-G0-G}(a), we show $\partial_{a}\Sigma_{\rm C}[G_{\rm C}]$ $(a=\omega, \varepsilon)$, which roughly counteract each other in Eq.~(\ref{eq:mass-tot}) between its denominator and numerator.
For comparison, we also present the results for $\Sigma_{\rm C}[G_{0}]$, which corresponds to the $G_{0}W_{0}$ approximation. The self-consistent iteration of the $GW_{0}$ formula simultaneously decreases the $\omega$-derivative~\cite{von_barth_self-consistent_1996} and increases the $\varepsilon$-derivative in absolute magnitude. The resulting total $m^{\ast}_{\rm el}$ is slightly smaller than unity, as shown in panel (b). These data are consistent with previous studies~\cite{hedin_new_1965,rietschel_role_1983}.

To clarify the roles of the $Z$ and $\chi$ terms, we also calculated the self-energy by iterating only with the antisymmetric part: $\Sigma_{\rm C}^{(Z)}[G_{\rm C}^{(Z)}]$.
The resulting $\partial_{\omega}\Sigma$ displayed in Fig.~\ref{fig:invmass-C-G0-G}(a) is reduced in magnitude compared to the first step ($G_{0}W_{0}$) but not as much as in the full self-energy case ($GW_{0}$).
This indicates that the $\chi$ term has a substantial effect in reducing $\partial_{\omega}\Sigma$.
Meanwhile, $\partial_{\varepsilon}\Sigma$, to which only the $\chi$ term contributes, increases through the iteration.
These behaviors can be understood as general characteristics of the self-energy.
The $Z$ term increases the effective mass and concomitantly decreases the spectral weight through the $\omega$-derivative.
Both are governed by the same factor, $1-\partial_{\omega}\Sigma$, appearing in Eq.~(\ref{eq:selfene_normal}) in a form $\xi_{i}/[1-\partial_{\omega}\Sigma]$ and a prefactor $[1-\partial_{\omega}\Sigma]^{-1}$, and thus the net number of states near the Fermi level ($k\simeq k_{\rm F}$ and $i\omega \simeq 0$) involved in the integral remains relatively unchanged.
On the other hand, including the $\chi$ term steepens the renormalized band, as indicated by the positive $\varepsilon$-derivative of $\partial_{\varepsilon}\Sigma$ in Fig.~\ref{fig:invmass-C-G0-G}(a).
The second iteration is thus performed with the effectively steeper quasiparticle band that hosts a reduced density of low-energy states.
Consequently, in the second iteration, both $\partial_{\omega}\Sigma$ and the {\it increment} of $\partial_{\varepsilon}\Sigma$ between first and second iteration steps are smaller in magnitude than in the first iteration, leading to convergence toward the $GW_{0}$ limit.

\subsubsection{Full self-energy case}

\begin{figure}[t!]
  \begin{center}
   \includegraphics[scale=0.60]{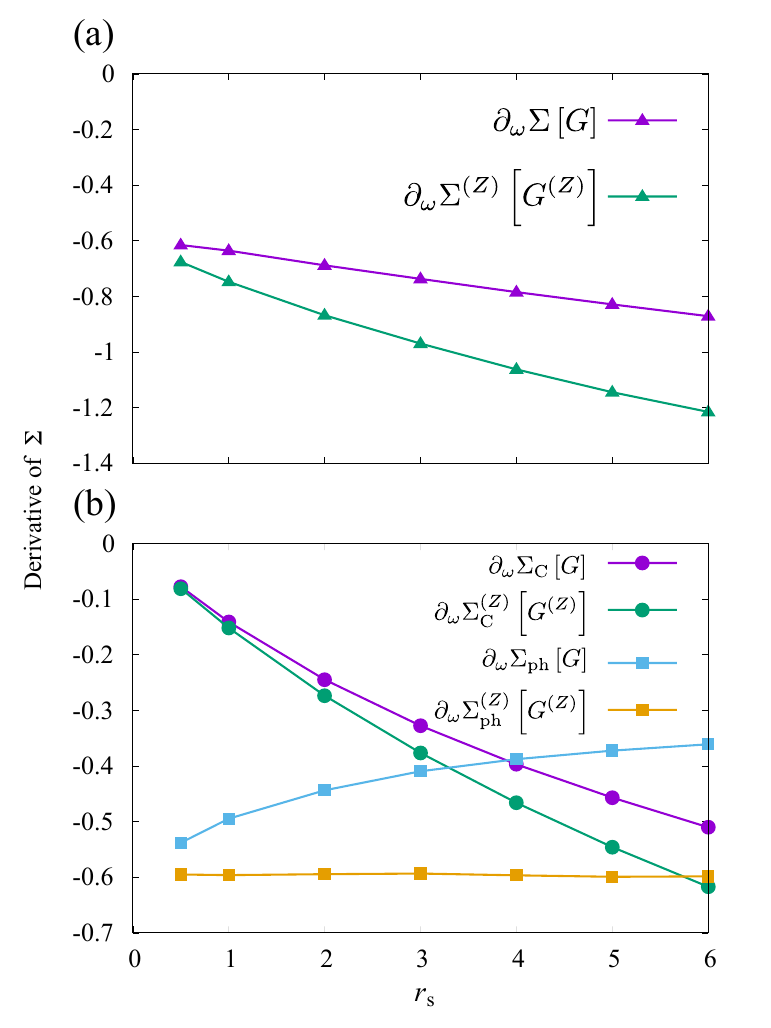}
   \caption{Normal self-energy properties calculated with both $W_{\rm ph}$ and $W_{\rm C}$. (a) The $\omega$-derivatives of the total self-energy converged with and without $\chi$ and (b) their decompositions into the phonic and Coulombic contributions.} 
   \label{fig:deriv_sigma_tot}
  \end{center}
\end{figure}

\begin{figure}[t!]
 \begin{center}
   \includegraphics[scale=0.60]{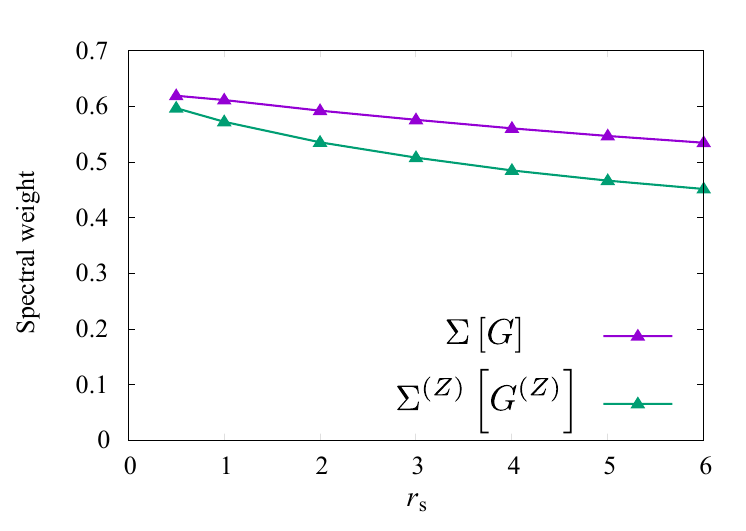}
   \caption{Spectral weight $z\equiv \left(1-\partial_{\omega}\Sigma_{\rm C}-\partial_{\omega}\Sigma_{\rm ph}\right)^{-1}$} 
   \label{fig:z_weight_tot}
 \end{center}
\end{figure}

The interplay of the $Z$ and $\chi$ terms through the self-consistent iteration originates from their general behaviors as revealed above. Indeed, we see similar reduction of $\partial_{\omega}\Sigma_{\rm ph}[G]$ (in absolute magnitude) if we execute the iteration for $G$ including the $\chi$ term. This was seemingly unrecognized in the literature since the separate treatment of $Z$ and $\chi$ was not motivated before the modern Eliashberg context and, as long as we apply the static approximation to the screened Coulomb interaction, $\chi$ is almost zero~\cite{Schrieffer-book}.

Figure \ref{fig:deriv_sigma_tot}(a) shows the $\omega$-derivative of $\Sigma\left[G\right]$ and $\Sigma^{(Z)}\left[G^{(Z)}\right]$ evaluated with $\lambda=0.6$, $\omega_{\rm ph}=300~K$, and $k_{\rm D}=2^{1/3}k_{\rm F}$. The decomposition into the phononic and Coulombic contributions are displayed in panel (b). $\partial_{\varepsilon}\Sigma_{\rm ph}$ was almost zero as anticipated and is not shown. It is well known that, in the Eliashberg theory, $\partial_{\omega}\Sigma_{\rm ph}^{(Z)}\left[G_{\rm ph}^{(Z)}\right]$ becomes approximately $-\lambda$, because of which the factor $1+\lambda$ is called mass renormalization factor. If the self-consistent iteration is executed with only the $Z$-term, the total $\omega$-derivative is approximately the sum of $\partial_{\omega}\Sigma_{\rm ph}^{(Z)}\left[G_{\rm ph}^{(Z)}\right]\simeq -\lambda$ and $\partial_{\omega}\Sigma_{\rm C}^{(Z)}\left[G_{\rm C}^{(Z)}\right]\simeq \partial_{\omega}\Sigma^{(Z)}_{C}\left[G^{(Z)}\right]$ (see also Fig.~\ref{fig:invmass-C-G0-G}~(a)). In contrast, if we execute the self-consistent iteration with the Coulombic $\chi$ term, both $\partial_{\omega}\Sigma_{\rm ph}$ and $\partial_{\omega}\Sigma_{\rm C}$ are reduced, resulting in far smaller total $\omega$-derivative. The total spectral weight at the Fermi level (Fig.~\ref{fig:z_weight_tot}) therefore becomes larger than expected from the simple sum of $\Sigma_{\rm ph}^{(Z)}\left[G_{\rm ph}^{(Z)}\right]+\Sigma_{\rm C}^{(Z)}\left[G_{\rm C}^{(Z)}\right]$. This is a synergy between electron-phonon and screened Coulomb interactions that would be missed with the static approximation to $W_{\rm C}$.

\begin{figure}[h!]
 \begin{center}
   \includegraphics[scale=0.65]{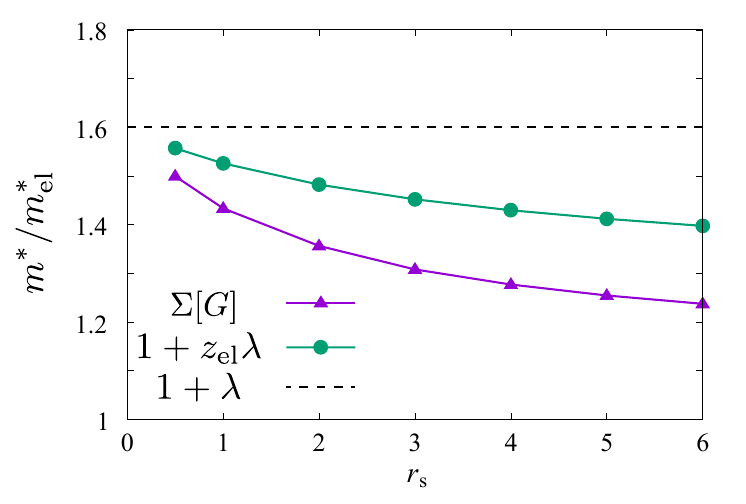}
   \caption{Effective mass renormalized by electron-phonon interaction, compared with the static Coulomb case $1+\lambda$ and Liu-Cohen formula $1+z_{\rm el}[G]\lambda$~\cite{liu_electron-phonon_1991}.} 
   \label{fig:effective_mass_elph}
 \end{center}
\end{figure}

The current analysis has important implications for experiments. The $\omega$-derivative of the phononic part is inferred from electronic low-energy properties like specific heat, assuming the mass renormalization factor $\simeq 1+\lambda$. But more precisely, the total effective mass is approximated as~\cite{liu_electron-phonon_1991}
\begin{eqnarray}
  m^{\ast}
  &\simeq& \frac{1-\partial_{\omega}\Sigma_{\rm C}}{1+\partial_{\varepsilon}\Sigma_{\rm C}}
  \times \left(1-\frac{\partial_{\omega}\Sigma_{\rm ph} }{1-\partial_{\omega}\Sigma_{\rm C}}\right)\nonumber \\
  &=&m^{\ast}_{\rm el}\left(1-z_{\rm el}\partial_{\omega}\Sigma_{\rm ph}\right)
\end{eqnarray}
with $z_{\rm el}\equiv\left(1-\partial_{\omega}\Sigma_{\rm C}\right)^{-1}$ being the spectral renormalization factor due to the Coulomb part. The correct mass enhancement due to the phonon-mediated attraction, $m^{\ast}/m^{\ast}_{\rm el}$, is therefore reduced by the Coulomb renormalization via $z_{\rm el}$ and $\partial_{\omega}\Sigma_{\rm ph}$. The former effect has been pointed out by Liu and Cohen~\cite{liu_electron-phonon_1991}, whereas we have clarified the latter: Whether we use $G$ or $G_{\rm ph}^{(Z)}$ for calculating $\partial_{\omega}\Sigma_{\rm ph}$ matters. The resulting factor $-z_{\rm el}\partial_{\omega}\Sigma_{\rm ph}$ becomes substantially smaller than $\lambda$, as shown in Fig.~\ref{fig:effective_mass_elph}.
When comparing the theory with experiments, this complication must be considered through a careful distinction in which model we implicitly define $\lambda$, before or after the Coulomb renormalization.

\begin{table*}[t!]
  \centering
  \caption{Superconducting $T_{\rm c}$ calculated with different approximations for self-energy.}
  \label{tab:tc_summary}
  \begin{tabularx}{420pt}{llll cccccc ccc}
    \toprule
    \multicolumn{4}{c}{Parameters}& \multicolumn{6}{c}{$T_{\rm c}$ (K) (Eliashberg)} & \multicolumn{3}{c}{$T_{\rm c}$ (K) (SCDFT)}\\
    \cmidrule(r){1-4}\cmidrule(r){5-10} \cmidrule(r){11-13}
     &&&& \multirow{2}{*}{$W_{\text{ph-only}}$} & \multirow{2}{*}{$W_{\rm Cstat}$}&\multicolumn{4}{c}{full $W_{\rm C}$}& && \\
     \cmidrule(r){7-10}
    $r_{\rm s}$ &$\lambda$ &$\omega_{\rm E}$(K) & $k_{\rm D}$($k_{\rm F}$) & & & $\Sigma_{\rm ph}[G_{\rm ph}]$ & $\Sigma[G_{0}]$ & $\Sigma^{(Z)}[G^{(Z)}]$ &$\Sigma[G]$ & AA2015 & LM2005 & SPG2020 \\
    \midrule
    \multirow{8}{*}{2.0} &\multirow{4}{*}{0.6} & \multirow{2}{*}{300} & $2^{1/3}$ &16.8&7.4 &17.8 &3.0  & 9.5 &4.4 &5.7& 3.4&7.6\\
                  && &    2.5  &17.1& 7.6& 18.1 & 3.1 & 9.7 &4.5  &5.7& 3.4&7.6\\
                  &  & \multirow{2}{*}{1000}  & $2^{1/3}$  &56.3&21.7& 58.7 & 9.9 & 31.3 &14.5 &15.4&9.2 &22.1\\
                  && & 2.5  &57.1&22.2& 59.6 &10.2  & 32.0 &14.9 &15.3& 9.1&22.1\\
                  &\multirow{4}{*}{1.2}  & \multirow{2}{*}{300}  & $2^{1/3}$  &41.6&29.3& 42.5 &14.7  & 30.8 &20.8 &22.8& 18.6&24.4 \\
                  && & 2.5  &41.9&29.6& 42.9 & 14.9 & 31.1 &21.1 &22.8&18.6 &24.4\\
                  &  & \multirow{2}{*}{1000}  & $2^{1/3}$  &139.0&93.5& 140.6& 48.9 & 102.1 &69.2 &66.6&56.4 &76.1\\
                  &&  & 2.5  &140.2&94.6& 141.8 & 49.7 & 103.2 &70.4 &66.3& 56.1&75.8\\
    \multirow{8}{*}{5.0} &\multirow{4}{*}{0.6} & \multirow{2}{*}{300} & $2^{1/3}$ &17.0&5.2 & 324.1 &6.9 & 48.8 &13.6  &4.4& 1.9&5.3\\
    &&& 2.5 &17.2&5.4 & 323.9 & 7.0& 49.0 &13.7 &4.4& 1.9&5.2\\
    && \multirow{2}{*}{1000}  & $2^{1/3}$ &57.7&13.9 & 402.6 & 18.0  & 109.5 &34.4 &9.1& 4.2&13.9\\
    &&& 2.5 &58.6&14.4 & 403.0 & 18.2& 110.1 &34.9 &8.8& 3.9&13.4\\
    &\multirow{4}{*}{1.2}  & \multirow{2}{*}{300}  & $2^{1/3}$ &41.9&26.0 & 308.0 &15.7  & 68.5 &29.6 &18.1& 14.3&20.3\\
    && & 2.5 &42.2&26.4  & 307.7 & 15.8 & 68.8 &29.8&18.0& 14.1&20.1\\
    & & \multirow{2}{*}{1000}  & $2^{1/3}$ &141.9&82.2 & 436.4 & 45.9 & 173.5 &85.2 &44.1&41.6 &62.2\\
    && & 2.5 &143.0&83.2 & 436.6 & 46.4 & 174.4 &85.9 &42.7& 39.7&59.8\\
    \bottomrule
  \end{tabularx}
\end{table*}

\begin{figure}[t!]
  \begin{center}
    \includegraphics[scale=0.60]{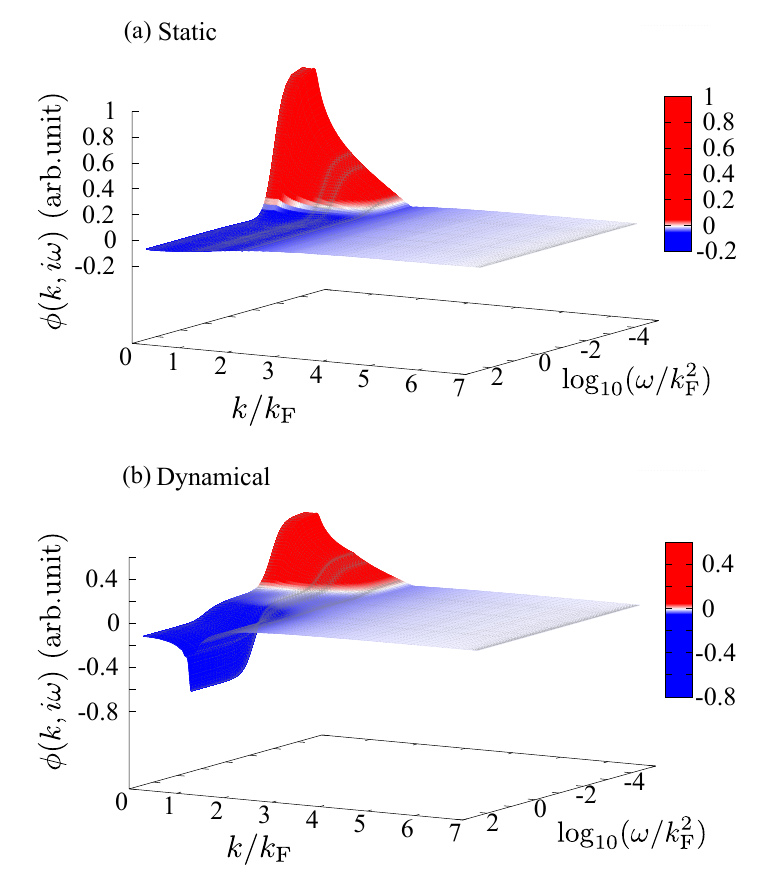}
    \caption{Normalized eigensolution of linearized Eliashberg equation $\phi(k,i\omega)$ at $T_{\rm c}$. The parameters $(r_{\rm s}, \lambda, \omega_{\rm E}, k_{\rm D})$ were set to $(2.0, 0.6, 300{\rm K}, 2^{1/3}k_{\rm F})$.
    The gray lines overlaid on the surfaces indicate the $k$ and $\omega$ grid points; concentration of non-uniform $k$ points at $0, k_{\rm F}, k_{\rm D}, 2k_{\rm F}$ is seen. (a) Results with static approximation to $W_{\rm C}$, and (b) full $W_{\rm C}$.} 
    \label{fig:Eliashberg-eigenvec}
  \end{center}
 \end{figure}

\subsection{Superconducting transition temperature}
We calculated superconducting $T_{\rm c}$ of the present model for varying model parameters $r_{\rm s}, \lambda, \omega_{\rm ph}, q_{\rm D}$: The former one controls the screening of the Coulomb interaction and absolute Fermi energy, whereas the latter three dominates the phononic properties for the model phonon-mediated attraction. The paramters were set so that it well represent realistic materials: $r_{\rm s}=2.0$ and $5.0$ are meant to represent elemental aluminum and alkali metals. $\lambda=0.6$ and $1.2$ respectively correspond to the weak- and strong-coupling regimes. $\omega_{\rm ph}$=300~K and 1000~K are for aluminum and compressed hydrides. The Debye wavenumber $k_{\rm D}$=$2^{1/3}k_{\rm F}$ corresponds to the situation where each atom provide one conducting electron to the medium, whereas $k_{\rm D}=2.5k_{\rm F}$ is intended to represent that the Umklapp scattering involves all the pair of wavenumbers at the Fermi surface: In the present case difference in $k_{\rm D}$ had little effect on $T_{\rm c}$, though.
We only calculated the $s$-wave solution to $\phi({\bf k},i\omega)\equiv\phi(k,i\omega)$.

For readers' reference, we display in Fig.~\ref{fig:Eliashberg-eigenvec} typical forms of the eigensolution of $\phi(k,i\omega)$ at $T_{\rm c}$ for cases of using the static and full $\omega$-dependent screened Coulomb interactions throughout the normal and anomalous calculations.
Generally, the anomalous self-energy eigensolution $\phi(k,i\omega)$ is positive in a tiny region $0\lesssim k \lesssim k_{\rm F}+k_{\rm D} \ \cap \ |\omega|\lesssim \omega_{\rm E}$ and becomes negative at larger $k$ and $\omega$, which indicate the retardation effect~\cite{morel_calculation_1962}.
The largest positive values are seen within $k<|k_{\rm F}-k_{\rm D}|$ but the small-wavenumber region is not so relevant as expected because of the denominator $\bar{\Theta}(k,i\omega)$ in Eq.~(\ref{eq:linear-Eliashberg}).
In the static case, the negative tail depends on $k$ and $\omega$ modestly, whereas in the dynamical case, drastic dependence is seen.
In particular, at $k\simeq k_{\rm F}$ we observe a steep valley at frequency $|\omega|\gtrsim k_{\rm F}^2$ originating from the long-range nature of the screened Coulomb interaction with nonzero frequency.
This structure is consistent with earlier results, which were derived through numerical extrapolation to low temperatures~\cite {takada_plasmon_1992,richardson_effective_1997}. 

The calculated $T_{\rm c}$ values are summarized in Table~\ref{tab:tc_summary}.
The results are classified into three, depending on the approximation for $W_{\rm C}$ used throughout the normal and anomalous self-energy calculations; ``$W_{\text{ph-only}}$'' for $W_{\rm C}=0$, ``$W_{\rm Cstat}$'' for $W_{\rm C}\simeq W_{\rm C}(i\nu=0)$ and ``full $W_{\rm C}$'' for full wavenumber-frequency dependent $W_{\rm C}$. The last case is further classified by the normal self-energy used for the anomalous self-energy calculation [Eq.~(\ref{eq:linear-Eliashberg})]. ``$\Sigma_{\rm ph}[G_{\rm ph}]$" denotes that the normal self-energy was calculated self-consistently with only $W_{\rm ph}$, as in the standard first-principles Eliashberg calculations with $W_{\rm Cstat}$. ``$\Sigma\left[G_{0}\right]$" denotes that we calculated $\Sigma_{\rm ph}+\Sigma_{\rm C}$ only once.
``$\Sigma^{(Z)}[G^{(Z)}]$" indicate that the normal self-energy was calculated self-consistently with full interaction $W_{\rm ph}+W_{\rm C}$ but through the antisymmetric formula (Eq.~(\ref{eq:selfene_antisym})). The last one ``$\Sigma[G]$" indicates the full self-consistent calculation with full interaction.

With the static Coulomb approximation (``$W_{\rm Cstat}$"), the Coulomb interaction only suppresses $T_{\rm c}$ from the phonon-only case (``$W_{\text{\rm ph-only}}$''). The magnitude of the suppression is larger in dilute ($r_{\rm s}=5.0$) regime due to poor screening. The effect of the normal self-energy was evaluated with four levels of approximation. If we do not include the normal self-energy from $W_{\rm C}$, ``$\Sigma_{\rm ph}[G_{\rm ph}]$", $T_{\rm c}$ is substantially larger than the full self-energy case, amounting to one order of magnitude larger. This behavior has been first pointed out by Rietschel and Sham~\cite{rietschel_role_1983} and is consistent with the results for real materials by Davydov {\it et al.}~\cite{davydov_ab_2020}. Including the normal self-energy $Z$ and $\chi$, this overestimation is alleviated [``$\Sigma[G]$"]. The $\Sigma[G_{0}]$ approximation suppresses $T_{\rm c}$~\cite{akashi_revisiting_2022} but too much than $\Sigma[G]$: Spectral weight $z$, that scales the integrand in Eq.~(\ref{eq:linear-Eliashberg}) by prefactor $z^2$, is partially recovered by the self-consistent iterations as shown in Fig.~\ref{fig:invmass-C-G0-G}(a) but with the $\Sigma[G_{0}]$ approximation this recovery does not occur. The $\Sigma^{(Z)}[G^{(Z)}]$ approximation~\cite{davydov_ab_2020} suppresses the overestimated $T_{\rm c}$, but only partially. The comparison between the $\Sigma^{(Z)}[G^{(Z)}]$ and $\Sigma[G]$ approximations shows us that the effect with $\chi$, that affects the pairing via the quasiparticle band steepening and the synergistic reduction of $\partial_{\omega}\Sigma$ shown above, is as significant as $Z$, as estimated previously~\cite{akashi_revisiting_2022}.

The positive effect of the dynamical structure has been proposed, for which one of the authors called plasmonic ``assistance'' in the phonon-mediated superconductors~\cite{akashi_development_2013}. Let us define this as the enhancement of $T_{\rm c}$ compared with the static Coulomb approximation. The full self-energy result shows that the assistance in this context does not become relevant in dense cases, or rather the plasmons suppress $T_{\rm c}$ further since the normal state renormalization dominates over the anomalous pairing contribution. The assistance is indeed active for $r_{\rm s}=5.0$, by which $T_{\rm c}$ becomes higher than $r_{\rm s}=2.0$ at odds with the static Coulomb case. Overall, the results indicate that if one wants to include the dynamical Coulomb effects, full inclusion of $Z$ and $\chi$ is mandatory in order to prevent unreasonable overestimation of $T_{\rm c}$.

\begin{table*}[t!]
  \centering
  \caption{Summary of the functional approximations for SCDFT.}
  \label{tab:SCDFT-funcs}
  \begin{tabular}{c c | ccc}
   \toprule
     Functional set& $W_{\rm C}$ feature &$K_{\rm ph}$& $Z_{\rm ph}$&$K_{\rm C}$ \\
    \hline
     LM2005& static &Eq.~(9) in Ref.~\cite{marques_ab_2005}&Eq.~(11) in Ref.~\cite{marques_ab_2005}&Eq.~(13) in Ref.~\cite{Massidda_2009}\\
     AA2015& dynamical&Eq.~(9) in Ref.~\cite{marques_ab_2005}&Eq.~(47) in Ref.~\cite{akashi_particle_hole_2013}&Eq.~(\ref{eq:K_C_dyn2})\\
     SPG2020& static&Eq.~(11) in Ref.~\cite{SPG_2020}&Eq.~(10) in Ref.~\cite{SPG_2020}&Eq.~(13) in Ref.~\cite{Massidda_2009}\\
    \bottomrule
  \end{tabular}
\end{table*}

\subsubsection*{Comparison with density functional theory for superconductors}
Finally, we calculate $T_{\rm c}$ of the same system with the density functional theory for superconductors (SCDFT)~\cite{oliveira_density-functional_1988,luders_ab_2005,marques_ab_2005}.
Similarly to the Kohn-Sham density functional theory~\cite{hohenberg_inhomogeneous_1964,kohn_self-consistent_1965,mermin_thermal_1965}, this theory is based on the theorem that the normal and anomalous electron densities of the interacting electron system at thermal equilibrium are calculated exactly by solving the effective one-particle equations, if the exact exchange-correlation potentials are employed.
This theory is tempting as it implies that the anomalous density, which characterizes the superconducting phase, can, in principle, be calculated exactly with the dynamical self-energy effect but without explicit dependence of the fundamental equation on frequencies. Exchange-correlation potentials describing the phonon-mediated pairing attraction and dynamical Coulomb interaction have been developed by referring to the Eliashberg theory. However, a comparison of the numerical behavior of the functionals to the Eliashberg theory with a dynamical screened Coulomb interaction, which we present below, was lacking.  

In the SCDFT, the superconducting phase is characterized by the linearized gap equation~\cite{luders_ab_2005,marques_ab_2005}
\begin{eqnarray}
  \Delta_{i}=-Z_{i}\Delta_{i}-\frac{1}{2}\sum_{i'}K_{ii'}\frac{{\rm tanh}\left(\beta\xi_{i'}/2\right)}{\xi_{i'}}\Delta_{i'}
  .
  \label{eq:SCDFT-eq}
\end{eqnarray}
Unlike the Eliashberg equations, this ``gap'' $\Delta$ is not directly related to the single-particle spectral gap, but still its onset exactly indicates the superconducting transition. The diagonal and non-diagonal kernels $Z_{i}$ and $K_{ii'}$ are functionals of the normal state electron and phonon properties. In this work we test three representative combinations of the functionals, in all of which the kernels are separable into the phononic and Coulombic contributions: $Z_{i}=Z_{{\rm ph},i}$ and $K_{ii'}=K_{{\rm ph},ii'}+K_{{\rm C},ii'}$. See Table~\ref{tab:SCDFT-funcs} for a summary.
The Coulombic diagonal term $Z_{{\rm C},i}$ including the dynamic screening has also been explored~\cite{davydov_ab_2020} but we do not include it in this work.
The local density approximation to $K_{{\rm C},ii'}$~\cite{kawamura_anisotropic_2017} becomes identical to the RPA in the current uniform system. 

We calculate $T_{\rm c}$ in the present model electron-phonon system with the SCDFT functional sets. Thanks to the spherical symmetry of the system, the gap equation for $s$-wave component $\Delta(p)$ simplifies to 
\begin{eqnarray}
  &&\bar{\Delta}(p)=-Z(p)\bar{\Delta}(p)\!-\!\int_{0}^{k_{\rm max}} \!\!\! dk \bar{K}(p,k) \frac{{\rm tanh}\left(\frac{\beta \xi(k)}{2}\right)}{2\xi(k)}\bar{\Delta}(k), \nonumber \\
  \\
  &&\bar{K}(p,k)=\frac{1}{4\pi^2}\int_{|p-k|}^{p+k} dq qK(q)
\end{eqnarray}
with $\bar{\Delta}(p)\equiv p\Delta(p)$.

LM2005~\cite{luders_ab_2005,marques_ab_2005} is composed of the phononic diagonal and non-diagonal terms, $Z_{\rm ph}$ and $K_{\rm ph}$, and Coulombic non-diagonal term $K_{\rm C}$. The former two kernels are formulated referring to the McMillan-Allen-Dynes equation~\cite{mcmillan_transition_1968,allen_transition_1975}. The Coulombic non-diagonal term is taken to be the Cooper matrix element of the screened Coulomb interaction
\begin{eqnarray}
K_{{\rm el},ii'}=W_{{\rm C},ii'}(\omega=0).
\end{eqnarray}
The original authors proposed that $W_{\rm C}$ be the Thomas-Fermi form, whereas in later applications $W_{\rm C}$ are often set to be of the static RPA~\cite{Massidda_2009}. We adopt the latter here.

Akashi and Arita proposed to include the $\omega$-dependence of $W_{\rm C}$ within the RPA~\cite{akashi_development_2013,akashi_density_2014,akashi_h2s_2015}. The kernel is formulated taking the $T\rightarrow 0$ limit as
\begin{eqnarray}
 && K_{{\rm C},ii'}=\lim_{T\rightarrow 0}\frac{1}{{\rm tanh}\left(\beta \xi_{i}/2\right){\rm tanh}\left(\beta \xi_{i'}/2\right)}\nonumber \\
 && \hspace{15pt} \times  \frac{1}{\beta^2}\sum_{\omega_{1}\omega_{2}}F_{i}(i\omega_{1})F_{i'}(i\omega_{2})W_{{\rm C},ii'}\left[i(\omega_{1}-\omega_{2})\right]
 \label{eq:K_C_dyn}
 \\
 && \hspace{15pt} =V_{{\rm C},ii'}+\frac{2}{\pi}\int_{0}^{\infty}dE\frac{{\rm Im}W_{{\rm C},ii'}(E+i0^{+})}{E+|\xi_{i}|+|\xi_{i'}|}
  \label{eq:K_C_dyn2}
\end{eqnarray}
with $F_{i}(i\omega)=\frac{1}{i\omega -\xi_{i}}-\frac{1}{i\omega +\xi_{i}}$. We find it has exactly the same form as the kernel in the Kirzhnits-Maximov-Khomskii equation~\cite{kirzhnits_description_1973,kawamura_anisotropic_2017}, which has been derived with a ``weak-coupling~\cite{takada_plasmon_1978}" approximation or $\phi_{i}(\omega)\propto \delta(\omega-\xi_{i})$ to the Eliashberg equations.
With this form, the plasmon-phonon cooperation mechanism may be treated, whereas a criticism has been raised that it may yield overestimation of $T_{\rm c}$~\cite{davydov_ab_2020}.
We label the combination of the functionals used in Ref.~\cite{akashi_h2s_2015} as AA2015, which is composed of the phononic kernels with slight modification from LM2005 for accuracy in strong asymmetry in the electronic DOS~\cite{akashi_particle_hole_2013} and $K_{\rm C}$ in Eq.~(\ref{eq:K_C_dyn}). 

Sanna, Pellegrini and Gross~\cite{SPG_2020} improved $K_{{\rm ph}}$ and $Z_{{\rm ph}}$ from LM2005 so that the $T_{\rm c}$ as well as anomalous density agree better with the Eliashberg equations for the case $W_{{\rm C}}=0$. We label the combination of their phononic kernels and the static-RPA form for $K_{\rm C}$ as SPG2020.

The calculated values of $T_{\rm c}$ with the three functional sets are compared with the Eliashberg results in Table~\ref{tab:tc_summary}. The AA2015 functional~\cite{akashi_h2s_2015}, which is made so that the plasmon effect is included, is found to yield very similar results with the full $\Sigma[G]$ case for $r_{\rm s}=2.0$, but a careful comparison reveals that this is due to a fortunate cancellation of two errors. First, the phononic kernels AA2015 are almost the same as LM2005 and have been found to underestimate the phonon-mediated attraction~\cite{SPG_2020}, which is also highlighted by the LM2005 results compared with $W_{\rm Cstat}$. On the other hand, the Coulomb kernel neglects the normal self-energy effects which would suppress $T_{\rm c}$. Their cancellation is the cause of the accuracy at $r_{\rm s}=2.0$. Going to the dilute regime $r_{\rm s}=5.0$, AA2015 gives smaller values than $\Sigma[G]$ due to the weak-coupling Kirzhnits-Maksimov-Khomskii approximation. SPG2020 is formulated so that the Eliashberg $W_{\text{ph-only}}$ result is reproduced~\cite{SPG_2020} and also retains practical accuracy for the static $W_{\rm C}$ case, as seen in comparison with $W_{\rm Cstat}$. But, when $\lambda$ is strong, we find SPG2020 tends to underestimate $T_{\rm c}$ than the Eliashberg calculations.
This underestimation becomes more severe for $\omega_{\rm ph}=1000$K.
Indeed this phenomenon was observed in the application to LaH$_{10}$~\cite{errea_quantum_2020}, which hosts large $\lambda$ and $\omega_{\rm ln}$. 
Thus, at the static Coulomb approximation, SPG2020 serves as a computationally cheaper counterpart to the Eliashberg equations, but with a subtle deficit in strong coupling systems with hard phonons.
For capturing the dynamical Coulomb effects, AA2015 is found to lack accuracy in plasmonic normal self-energy and strong coupling effects.
It can nevertheless yield accurate values of $T_{\rm c}$ in dense systems, but care must be taken when analyzing the superconducting mechanism. 

\section{Summary}
Despite recent rapid computational advances, the numerical behavior of the Eliashberg equations with the frequency-dependent screened Coulomb interaction was not well understood because of the slow convergence in wavenumber and frequency cutoff and the requirement of reaching extremely low temperatures. We have overcome those obstacles by the use of the intermediate basis representation, the Fourier convolution theorem, and the continuous Euler quadrature. Applying our method to the uniform electron gas model, we have quantitatively established how the dynamical structure of the screened Coulomb interaction cancels or cooperates with the phonon-mediated pairing attraction in practical $r_{\rm s}$ settings. Our simple model and $GW_{0}$-Eliashberg results for it serve a reliable benchmark to which the first-principles superconducting methods may be compared, as demonstrated above for SCDFT. The numerical methods presented here may also be utilized for further extension for the vertex correction effects included as the local-field factors~\cite{kukkonen_electron-electron_1979,vignale_effective_1985,richardson_effective_1997,pellegrini_ab_2023,wang_origin_2023}. Extensions to real frequency and non-uniform systems are to be attempted.

\acknowledgments
This work was supported by JSPS KAKENHI Grant No. 23H03817 from the Japan Society for the Promotion of Science (JSPS).
This work utilized AI-based tools solely for language editing, such as grammar and spell checking.

\bibliography{reference}

\end{document}